\documentclass{emulateapj}

\usepackage{amssymb}

\def\ts     {\thinspace}
\def\kms    {\ts km\ts s$^{-1}$}
\def\etal   {{\rm et\ts al.}}
\def\msol   {$M_{\odot}$}

\def\lsol   {$L_{\odot}$}
\def\zsol   {$Z_{\odot}$}

\def\lprime {K\,\ts km\ts s$^{-1}$\,pc$^2$}

\def\aco    {{\rm CO}($J$=1$\to$0)}

\def\cco    {{\rm CO}($J$=3$\to$2)}

\slugcomment{draft version \today, accepted for publication in the Astrophysical Journal Letters}

\shorttitle{\aco\ Observations of $z$$\sim$3 Lyman-Break Galaxies}
\shortauthors{Riechers et al.}

\begin{document}

\title{Total Molecular Gas Masses of $z$$\sim$3 Lyman-Break Galaxies:\\
 \aco\ Emission in MS\,1512-cB58 and the Cosmic Eye}

\author{Dominik A.\ Riechers\altaffilmark{1,4}, Christopher L.\ Carilli\altaffilmark{2}, Fabian Walter\altaffilmark{3}, and Emmanuel Momjian\altaffilmark{2}}

\altaffiltext{1}{Astronomy Department, California Institute of
  Technology, MC 249-17, 1200 East California Boulevard, Pasadena, CA
  91125, USA; dr@caltech.edu}

\altaffiltext{2}{National Radio Astronomy Observatory, PO Box O, Socorro, NM 87801, USA}

\altaffiltext{3}{Max-Planck-Institut f\"ur Astronomie, K\"onigstuhl 17, D-69117 Heidelberg, Germany}

\altaffiltext{4}{Hubble Fellow}


\begin{abstract}

We report the detection of \aco\ emission toward the lensed
$L^{\star}_{\rm UV}$ Lyman-break galaxies (LBGs) MS\,1512-cB58
($z$=2.73) and the Cosmic Eye ($z$=3.07), using the Expanded Very
Large Array. The strength of the CO line emission reveals molecular
gas reservoirs with masses of (4.6$\pm$1.1)$\times$10$^8$\,$(\mu_{\rm
L}/32)$$^{-1}$\,$(\alpha_{\rm CO}/0.8)$\,\msol\ and
(9.3$\pm$1.6)$\times$10$^8$\,$(\mu_{\rm L}/28)$$^{-1}$\,$(\alpha_{\rm
CO}/0.8)$\,\msol, respectively. These observations suggest by
$\sim$30\%--40\% larger gas reservoirs than estimated previously based on
\cco\ observations due to subthermal excitation of the $J$=3
line. These observations also suggest gas mass fractions of
0.46$\pm$0.17 and 0.16$\pm$0.06. The \aco\ emission in the Cosmic Eye
is slightly resolved on scales of 4.5$''$$\pm$1.5$''$, consistent with
previous studies of nebular emission lines. This suggests that the
molecular gas is associated with the most intensely star-forming
regions seen in the ultraviolet (UV). We do not resolve the \aco\
emission in cB58 at $\sim$2$''$ resolution, but find that the \aco\
emission is also consistent with the position of the UV-brightest
emission peak. The gas masses, gas fractions, moderate CO line
excitation, and star formation efficiencies in these galaxies are
consistent with what is found in nearby luminous infrared
galaxies. These observations thus currently represent the best
constraints on the molecular gas content of `ordinary' (i.e.,
$\sim$$L^{\star}_{\rm UV}$) $z$$\sim$3 star-forming galaxies. Despite
comparable star formation rates, the gas properties of these young
LBGs seem to be different from the recently identified
optical/infrared-selected high-$z$ massive, gas-rich star-forming
galaxies, which are more gas-rich and massive, but have lower star
formation efficiencies, and presumably trace a different galaxy
population.

\end{abstract}

\keywords{galaxies: active --- galaxies: starburst --- 
galaxies: formation --- galaxies: high-redshift --- cosmology:
observations --- radio lines: galaxies}

\section{Introduction}

A substantial fraction of the star formation rate density (SFRD) of
the universe at $z$$>$2 occurs in young star-forming galaxies that can
be identified through a significant Lyman break in their spectra,
so-called Lyman-break galaxies (LBGs; e.g., Steidel et al.\
\citeyear{ste96}). It was recently suggested that $>$25\% of the
stellar mass in the universe was formed in LBGs (Reddy \& Steidel
\citeyear{rs09}). These galaxies typically have UV-derived star 
formation rates (SFRs) in excess of 10\,\msol\,yr$^{-1}$ (e.g., Nandra
et al.\ \citeyear{nan02}; Reddy et al.\ \citeyear{red10}), marking
comparatively moderate star formation events in view of the $\sim$2
orders of magnitude higher SFRs usually found in distant submillimeter
galaxies (SMGs) and quasars (e.g., Blain et al.\ \citeyear{bla02};
Wang et al.\ \citeyear{wan08}). However, due to the presence of dust
in the star-forming regions, their exact properties (and thus, the
absolute contribution to the SFRD) are difficult to determine (e.g.,
Reddy et al.\ \citeyear{red06}; Wilkins et al.\ \citeyear{wil08};
Carilli et al.\ \citeyear{car08}; Siana et al.\ \citeyear{sia08}). It
is thus desirable to study the gas and dust properties of LBGs
directly, in particular through molecular line emission.

Due to the low (stellar) masses and gas content of LBGs (e.g., Shapley
et al.\ \citeyear{sha03}; compared to other systems observed at high
redshift, e.g., SMGs and quasars), such studies are currently only
possible in strongly lensed systems.  Due to lensing magnification
factors of $\sim$30, two $z$$\sim$3 lensed LBGs were successfully
detected in \cco\ emission (Baker et al.\ \citeyear{bak04}; Coppin et
al.\ \citeyear{cop07}). These observations revealed the presence of
substantial amounts of molecular gas. However, recent studies of the
ground-state \aco\ transition in high-$z$ galaxies have shown that
this line appears to carry a higher luminosity than higher-$J$ CO
lines in many cases, and that the ratio appears to be a function of
galaxy type (e.g., Riechers et al.\ \citeyear{rie06},
\citeyear{rie10}; Hainline et al.\ \citeyear{hai06}; Dannerbauer et
al.\ \citeyear{dan09}; Ivison et al.\ \citeyear{ivi10}; Carilli et
al.\ \citeyear{car10}; Harris et al.\ \citeyear{har10}; Aravena et
al.\ \citeyear{ara10}). Thus, \aco\ observations are crucial to
determine the total molecular gas content in high-$z$ galaxies,
independent of gas excitation conditions. To address this issue for
LBGs, we have targeted both LBGs previously detected in
\cco\ emission in the \aco\ line.

In this letter we report the detection of \aco\ emission toward the
strongly lensed $L^{\star}_{\rm UV}$ $z$$\sim$3 LBGs MS\,1512-cB58
(Yee et al.\ \citeyear{yee96}; Pettini et al.\ \citeyear{pet00},
\citeyear{pet02}) and the Cosmic Eye (Smail et al.\ \citeyear{sma07}),
using the Expanded Very Large Array (EVLA). We use a concordance, flat
$\Lambda$CDM cosmology throughout, with $H_0$=71\,\kms\,Mpc$^{-1}$,
$\Omega_{\rm M}$=0.27, and $\Omega_{\Lambda}$=0.73 (Spergel \etal\
\citeyear{spe03}, \citeyear{spe07}).

\section{Observations}

\begin{figure*}
\vspace{-5mm}

\epsscale{1.15}
\plotone{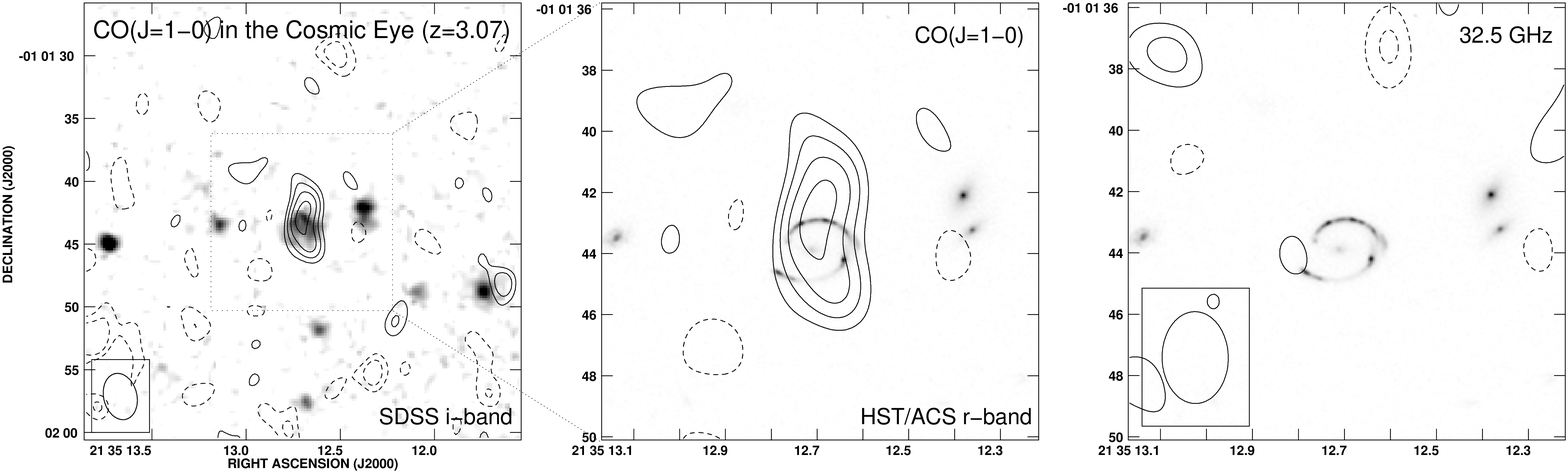}
\vspace{-7mm}

\caption{EVLA contour maps of \aco\ emission toward the Cosmic Eye, 
integrated over the central 200\,\kms\ (18.75\,MHz). {\em Left}:\
Emission overlayed on a grey-scale image of the $i$-band continuum
emission (from the Sloan Digital Sky Survey). Contours are shown in
steps of 1$\sigma$=45\,$\mu$Jy\,beam$^{-1}$, starting at
$\pm$2$\sigma$. The beam size of 3.4$''$$\times$2.6$''$ is shown in
the bottom left. {\em Middle}:\ Same contours, but zoomed-in and
overlayed on a high-resolution {\em HST}/ACS F606W image (Smail et
al.\ \citeyear{sma07}). {\rm Right}:\ Same, but with 32.5\,GHz
contours overlayed (beam size is 3.0$''$$\times$2.2$''$; same contour
levels in all panels). No 9.2\,mm continuum emission is detected down
to a 2$\sigma$ limit of 72\,$\mu$Jy\,beam$^{-1}$. \label{f1}}
%
\end{figure*}

We observed the \aco\ ($\nu_{\rm rest} = 115.2712$\,GHz) emission line
toward the Cosmic Eye and MS\,1512-cB58 (hereafter:\ cB58) using the
EVLA.  At $z$=3.074 and 2.727, this line is redshifted to 28.2944 and
30.9287\,GHz (10.6 and 9.7\,mm), respectively. Observations were
carried out under good 9\,mm weather conditions during 4\,tracks in D
array between 2009 December 4 and 28, resulting in 8.2 and 7.3\,hr
on-source time with 16 and 15\,antennas after rejection of bad
data. The nearby quasars J2134-0153 (distance to the Cosmic Eye:\
$0.5^\circ$) and J1506+3730 (distance to cB58:\ $1.5^\circ$) were
observed every 3.5\,minutes for pointing, secondary amplitude and
phase calibration. For primary flux calibration, the standard
calibrators 3C48 and 3C286 were observed, leading to a calibration
that is accurate within $\lesssim$10\%.

Observations were carried out with the previous generation correlator
and set up using two intermediate frequencies (IFs) with a bandwidth
of 21.875\,MHz ($\sim$220\,\kms, dual polarization) each, and a
resolution of 3.125\,MHz ($\sim$32\,\kms ). Due to tuning
restrictions, we centered the first IF on the \aco\ line, and the
second IF at 32.5\,GHz to measure limits on the continuum emission
close to the line.

For data reduction and analysis, the AIPS package was used.  All data
were mapped using `natural' weighting unless mentioned otherwise. For
the Cosmic Eye, the data result in a final rms of
70/45\,$\mu$Jy\,beam$^{-1}$ per 100/200\,\kms\ (9.375/18.75\,MHz)
channel at a synthesized clean beam size of 3.4$''$$\times$2.6$''$
(3.0$''$$\times$2.2$''$ at 32.5\,GHz). For MS1512-cB58, the data
result in a final rms of 70\,$\mu$Jy\,beam$^{-1}$ per 120\,\kms\
(12.5\,MHz) channel at a synthesized clean beam size of
2.9$''$$\times$2.4$''$ (`robust 0' weighting:\ 2.1$''$$\times$1.8$''$
beam; 1.9$''$$\times$1.6$''$ at 32.5\,GHz).

\begin{figure*}
\vspace{-9mm}

\epsscale{1.15}
\plotone{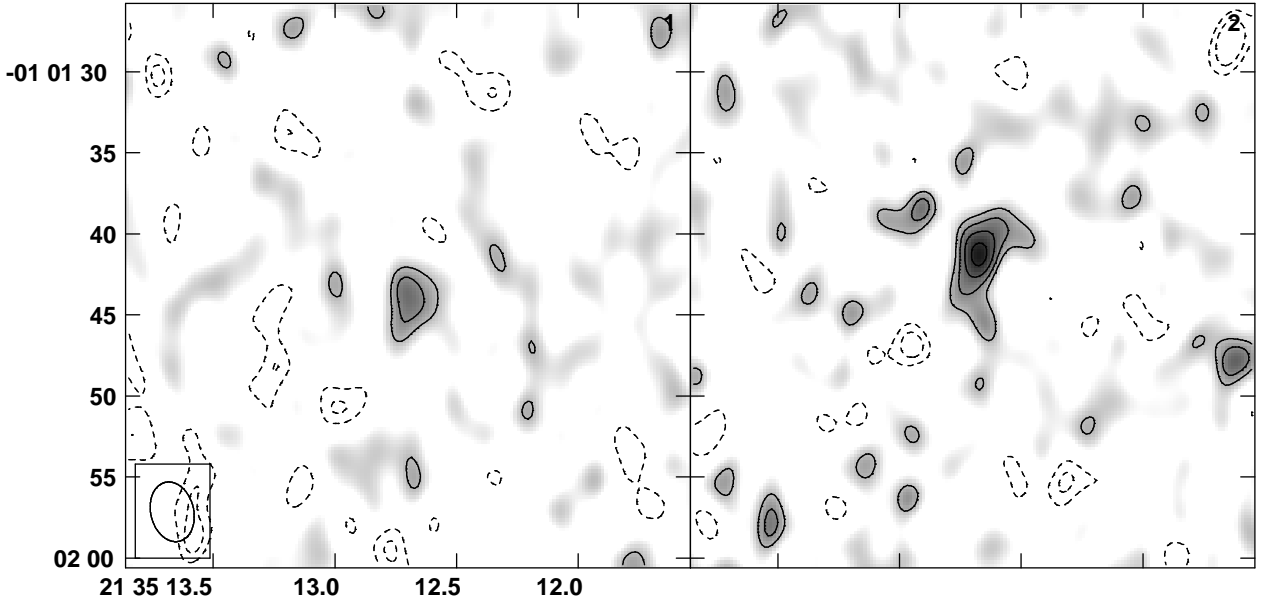}
\vspace{-16mm}

\caption{Red/blue channel maps of the \aco\ emission in the Cosmic Eye. The 
same region is shown as in Fig.~\ref{f1} ({\em left}). One channel
width is 100\,\kms\ (9.375\,MHz). Contours are shown in steps of
1$\sigma$=70\,$\mu$Jy\,beam$^{-1}$, starting at $\pm$2$\sigma$. The
beam size is shown in the bottom left.
\label{f2}}
%
\end{figure*}

\section{Results}

\subsection{The Cosmic Eye}

We have detected spatially resolved \aco\ line emission toward the
$z$=3.07 Cosmic Eye at $\sim$6.5$\sigma$ significance (Fig.~\ref{f1}).
From two-dimensional Gaussian fitting, we measure a deconvolved source
size of 4.5$''$$\pm$1.5$''$ along the north-south axis, consistent
with the UV continuum size within the errors. The source is unresolved
along the east-west axis down to $\simeq$2$''$. Given the relative
errors, this is also consistent with the upper limit on the size of
the \cco\ emission of $\lesssim$3$''$ (Coppin et al.\
\citeyear{cop07}). The peak of the \aco\ emission (peak position:\
21$^{\rm h}$35$^{\rm m}$12.$^{\rm s}$700$\pm$0.$^{\rm s}$013;
--01$^\circ$01$'$42.$''$70$\pm$0.$''$47) is consistent with the
brightest emission region along the northern lens arc of this LBG at
606\,nm ({\em HST} $r$-band; rest-frame 149\,nm) and 762.5\,nm (SDSS
$i$-band; rest-frame 187\,nm; see Fig.~\ref{f1}). We thus conclude
that the CO line and UV continuum emission emerge from the same
star-forming regions (leading us to adopt the same lensing
magnification), contrary to previous suggestions by Coppin et al.\
(\citeyear{cop07}) based on \cco\ measurements (their peak position:\
21$^{\rm h}$35$^{\rm m}$12.$^{\rm s}$62; --01$^\circ$01$'$43.$''$9;
see Sect.\ 3.1 in Coppin et al.\ \citeyear{cop07}).  No 9.2\,mm
continuum emission is detected down to a 2$\sigma$ limit of
72\,$\mu$Jy\,beam$^{-1}$ (Fig.~\ref{f1}; {\em right} panel).

In Figure~\ref{f2}, the \aco\ emission is shown in two 100\,\kms\ wide,
red and blue velocity channels. Within the limited signal--to--noise
ratio, the channel maps indicate that the emission is not only
spatially, but also dynamically resolved, with molecular gas moving
from north to south between the blue and red line wings. Due to the
source's complex lensing morphology, the direction of the velocity
gradient cannot simply be translated into a rotation axis in the
source plane. Higher resolution and signal--to--noise observations
with the full EVLA are required to investigate the dynamical structure
of the gas reservoir in more detail.

From the Gaussian fit, we measure a line peak strength of
262$\pm$45\,$\mu$Jy\,beam$^{-1}$, and a spatially-integrated strength
of 365$\pm$107\,$\mu$Jy (over 200\,\kms, comparable to the CO
$J$=3$\to$2 line FWHM of 190\,\kms; Coppin et al.\
\citeyear{cop07}). This corresponds to a \aco\ line intensity of
$I_{\rm CO}$=0.077$\pm$0.013\,Jy\,\kms, and a line luminosity of
$L'_{\rm CO(1-0)}$=(3.27$\pm$0.56)$\times$10$^{10}$\,$\mu_{\rm
L}^{-1}$\,\lprime\ (where $\mu_{\rm L}$ is the lensing magnification
factor). We also derive a CO $J$=3$\to$2/1$\to$0 line brightness
temperature ratio\footnote{We here assume that the \aco\ and \cco\
emission emerge from the same gas component, and thus are lensed by
the same $\mu_{\rm L}$.} of $r_{31}$=0.72$\pm$0.16, suggesting that
the \cco\ line is subthermally excited.

\subsection{MS\,1512-cB58}

\begin{figure*}
\vspace{-5mm}

\epsscale{1.15}
\plotone{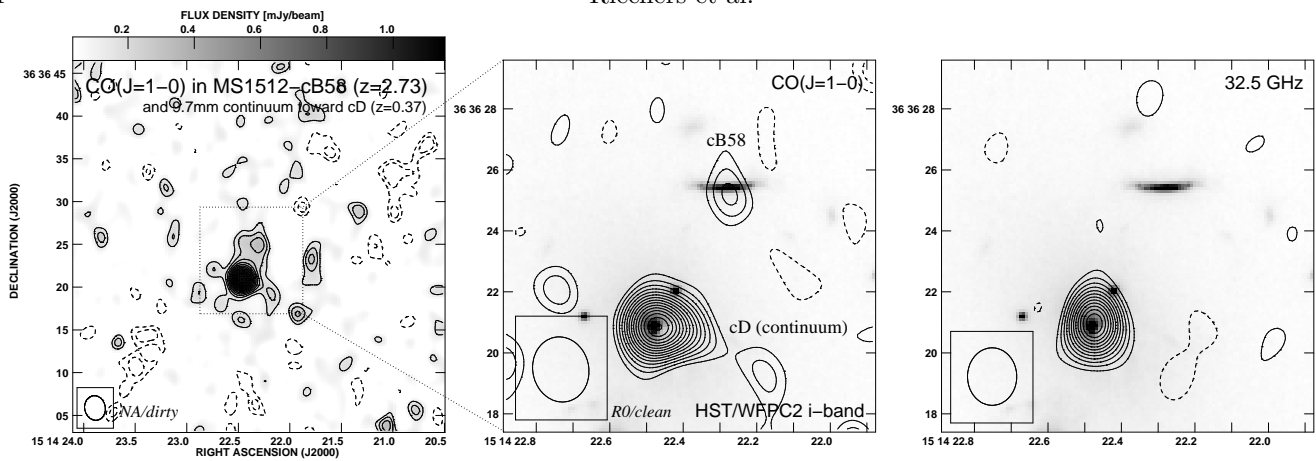}
\vspace{-9mm}

\caption{Contour maps of \aco\ emission toward MS\,1512-cB58, integrated over
the central 120\,\kms\ (12.5\,MHz). {\em Left}:\ Dirty image of the
emission, imaged using `natural' weighting. The second, brighter
source is radio continuum emission from the cD galaxy of the
foreground lensing cluster. Contours are shown in steps of
1$\sigma$=70\,$\mu$Jy\,beam$^{-1}$, starting at $\pm$2$\sigma$. The
beam size of 2.9$''$$\times$2.4$''$ is shown in the bottom left. {\em
Middle}: Cleaned image of the same, but zoomed-in, imaged using
`robust 0' weighting, and overlayed on a high-resolution {\em
HST}/WFPC2 F814W image (image from the Hubble Legacy Archive). The
beam size of 2.1$''$$\times$1.8$''$ is shown in the bottom left. {\rm
Right}:\ Same, but with 32.5\,GHz contours overlayed (beam size is
1.9$''$$\times$1.6$''$; same contour levels in all panels). No 9.2\,mm
continuum emission is detected toward cB58 down to a 2$\sigma$ limit
of 96\,$\mu$Jy\,beam$^{-1}$.
\label{f3}}
%
\end{figure*}

We have detected \aco\ emission toward the main lens arc of the
$z$=2.73 LBG MS\,1512-cB58 at $\sim$4.5$\sigma$ significance
(Fig.~\ref{f3}).\footnote{We do not detect the $>$10$\times$ less
magnified counterimage (Seitz et al.\ \citeyear{sei98}), as expected.}
Within $\sim$4.5$''$ of cB58, we detect a second, several times
brighter source (0.98$\pm$0.05 and 1.21$\pm$0.07\,mJy at 9.2 and
9.7\,mm), which we interpret to be continuum emission from the cD
galaxy in the lensing cluster in front of cB58 at $z$=0.37 (see gray
scale 814\,nm image in the {\em middle} panel of Fig.~\ref{f3}). This
separation is only $\sim$1.5$\times$ the resolution along the
separation axis (imaging the data with `natural' weighting), which is
sufficient for identification of both sources, but results in some
flux contribution of the cD galaxy at the position of cB58 (due to
sidelobe structure in the synthesized beam; see `dirty' map in the
{\em left} panel of Fig.~\ref{f3}). To properly separate the flux from
both sources, we thus imaged the emission with `robust 0' baseline
weighting (yielding higher spatial resolution) before applying the
CLEAN algorithm (Fig.~\ref{f3}; {\em middle} panel). Within the
errors, the peak of the \aco\ emission is consistent with the peak of
the 814\,nm ({\em HST} $i$-band; rest-frame 218\,nm) continuum
emission (Fig.~\ref{f3}; {\em middle} panel). No 9.2\,mm continuum
emission is detected toward cB58 down to a 2$\sigma$ limit of
96\,$\mu$Jy\,beam$^{-1}$ (Fig.~\ref{f3}; {\em right} panel).

After deconvolution using `robust 0' weighting, we measure a \aco\
line peak strength of 285$\pm$70\,$\mu$Jy for cB58. This corresponds
to $I_{\rm CO}$=0.052$\pm$0.013\,Jy\,\kms\ (assuming a line FWHM of
174\,\kms\ as for the CO $J$=3$\to$2 line; Baker et al.\
\citeyear{bak04}), and $L'_{\rm
CO(1-0)}$=(1.82$\pm$0.45)$\times$10$^{10}$\,$\mu_{\rm
L}^{-1}$\,\lprime.  We find $r_{31}$=0.78$\pm$0.25, similar to what is
found for the Cosmic Eye.  We also set an upper limit of
$r_{71}$$<$0.25 (3$\sigma$) on the CO $J$=7$\to$6/1$\to$0 ratio.

\section{Analysis and Discussion}

\subsection{Total Molecular Gas Masses}

Gas masses in LBGs are typically constrained from their H$\alpha$
luminosity, converting the (extinction-corrected) H$\alpha$ flux into
a SFR, and then using the star formation law (e.g., Kennicutt
\citeyear{ken98}) to convert the SFR to molecular gas mass ($M_{\rm
gas}$; e.g., Erb et al.\ \citeyear{erb06}). Besides its intrinsic
scatter, the star formation law implicitly depends on a conversion
factor from $L'_{\rm CO}$ to $M_{\rm gas}$ ($\alpha_{\rm CO}$),
yielding (at least) four considerable sources of uncertainty in such
estimates.

Gas mass estimates based on \aco\ depend on $\alpha_{\rm CO}$, but are
independent of other sources of uncertainty that are inherent to
alternative estimators. Thus, \aco\ is the best known diagnostic to
constrain total molecular gas masses in galaxies. Motivated by our
findings below, we here adopt $\alpha_{\rm
CO}$=0.8\,\msol\,(\lprime)$^{-1}$, as found in nearby luminous and
ultra-luminous infrared galaxies ((U)LIRGs; Downes \& Solomon
\citeyear{ds98}), rather than the higher values suggested for spirals
(see, e.g., Solomon \& Vanden Bout
\citeyear{sv05}).  Adopting $\mu_{\rm L}$=28 (Dye et al.\
\citeyear{dye07}) and $\mu_{\rm L}$=32 (Seitz et al.\
\citeyear{sei98}; Baker et al.\ \citeyear{bak04}), this yields $M_{\rm
gas}$=(9.3$\pm$1.6)$\times$10$^8$\,$(\mu_{\rm
L}/28)$$^{-1}$\,$(\alpha_{\rm CO}/0.8)$\,\msol\ and $M_{\rm
gas}$=(4.6$\pm$1.1)$\times$10$^8$\,$(\mu_{\rm
L}/32)$$^{-1}$\,$(\alpha_{\rm CO}/0.8)$\,\msol\ for the Cosmic Eye and
cB58, respectively.

\subsection{Specific Star Formation Rates and Mass Doubling Timescales}

Due to the presence of substantial amounts of dust and the typically
young age of LBGs, estimates of SFRs and stellar masses ($M_\star$)
usually agree only within a factor of a few between different
estimators and fits to the spectral energy distribution (e.g., Carilli
et al.\ \citeyear{car08}; Siana et al.\
\citeyear{sia08}). For cB58, we adopt a SFR of (25$\pm$10)\,\msol\,yr$^{-1}$ 
and $M_\star$=(1.0$\pm$0.3)$\times$10$^9$\,\msol, as well as an
infrared luminosity of $L_{\rm
IR}$=(1.5$\pm$0.8)$\times$10$^{11}$\,\lsol\ (Baker et al.\
\citeyear{bak04}; Siana et al.\ \citeyear{sia08}; see also Pettini et 
al.\ \citeyear{pet00}, \citeyear{pet02}). For the Cosmic Eye, we adopt
a SFR of 140$\pm$80\,\msol\,yr$^{-1}$,
$M_\star$=(6$\pm$2)$\times$10$^9$\,\msol, and $L_{\rm
IR}$=(8.3$\pm$4.4)$\times$10$^{11}$\,\lsol\ (Coppin et al.\
\citeyear{cop07}; Siana et al.\ \citeyear{sia09}).

These literature values yield specific star formation rates (SSFRs;
i.e., SFR/$M_\star$) of 25$\pm$12\,Gyr$^{-1}$ and
23$\pm$15\,Gyr$^{-1}$ and stellar mass doubling timescales of
$\tau_{\rm double}$=40$\pm$20\,Myr and 43$\pm$28\,Myr for cB58 and the
Cosmic Eye, respectively. Despite the fact that both sources are
$\sim$$L^{\star}_{\rm UV}$ LBGs at $z$$\sim$3, their SSFRs are
$\sim$5$\times$ higher and their $\tau_{\rm double}$ are
$\sim$5$\times$ lower than the median values for $z$$\sim$3 LBGs
(4.3\,Gyr$^{-1}$ and 230\,Myr; e.g., Magdis et al.\ \citeyear{mag10}).
We estimate that both the SSFR and $\tau_{\rm double}$ in cB58 and the
Cosmic Eye are by a factor of $\sim$3 uncertain (in particular due to
the difficulty in constraining any old part of their stellar
populations; Siana et al.\ \citeyear{sia08}), and thus, may well fall
within the scatter of the values found for the general (unlensed) LBG
population. If taken at face value, this could also imply that cB58
and the Cosmic Eye are close to the peak intensity of the starbursts
that drive the buildup of their stellar mass, when their SSFRs may
reach levels comparable to those in $z$$>$2 SMGs (15--30\,Gyr$^{-1}$;
e.g., Daddi et al.\ \citeyear{dad09}; Tacconi et al.\
\citeyear{tac08}). Indeed, both cB58 and the Cosmic Eye appear to be
comparatively young LBGs ($<$300\,Myr; e.g., Siana et al.\
\citeyear{sia08}, \citeyear{sia09}).

\subsection{Gas Fractions, Depletion Timescales, and Star Formation Efficiencies}

Both of our targets are gas-rich. We find gas mass
fractions\footnote{An index 0.8 indicates that $\alpha_{\rm
CO}$=0.8\,\msol\,(\lprime)$^{-1}$ is assumed.}$^,$\footnote{We do not
derive $M_{\rm dyn}$-based gas fractions due to the limited
constraints on the CO sizes and dynamics.} of $f_{\rm
gas}^{0.8}$=$M_{\rm gas}$/$M_\star$=0.46$\pm$0.17 and 0.16$\pm$0.06
and baryonic gas mass fractions of $f_{\rm bary}^{\rm g,0.8}$=$M_{\rm
gas}$/($M_{\rm gas}$+$M_\star$)=0.32$\pm$0.08 and 0.13$\pm$0.04 for
cB58 and the Cosmic Eye, respectively. These values are comparable to
nearby luminous and ultra-luminous infrared galaxies, but (on average)
somewhat lower than in SMGs (typical $f_{\rm bary}^{\rm g}$$\sim$0.4;
Tacconi et al.\ \citeyear{tac06}, \citeyear{tac08}) and high-$z$
massive, gas-rich star-forming galaxies (hereafter:\
SFGs;\footnote{These are galaxies with SFRs of $>$50\,\msol\,yr$^{-1}$
and $M_\star$$>$3$\times$10$^{10}$\,\msol\ selected in the
UV/optical/near-infrared; referred to in the literature as, e.g.,
`BzK' galaxies, `BX/BM'/AEGIS galaxies, or `normal' high-$z$
star-forming galaxies (e.g., Daddi et al.\ \citeyear{dad10}; Tacconi
et al.\ \citeyear{tac10}).} typical $f_{\rm bary}^{\rm
g}$$\sim$0.45--0.6; e.g., Daddi et al.\ \citeyear{dad08},
\citeyear{dad10}; Tacconi et al.\ \citeyear{tac10}). The comparatively
high $f_{\rm gas}$ and $f_{\rm bary}^{\rm g}$ in cB58 are also
consistent with its relatively young age ($<$30\,Myr; Siana et al.\
\citeyear{sia08}), and thus, a relatively early phase in its
starburst. On the other hand, cB58 may have had a higher SFR in the
past if all of its estimated stellar mass were build up in the ongoing
starburst within $<$30\,Myr.

The minimum times for which the starbursts can be maintained at their
current rates are given by the gas depletion timescales, which we find
to be $\tau_{\rm dep}^{0.8}$=$M_{\rm gas}$/SFR$\sim$18$\pm$8\,Myr and
$\sim$7$\pm$4\,Myr for cB58 and the Cosmic Eye,
respectively.\footnote{The main differences relative to previous
values are the different adopted $\alpha_{\rm CO}$ and a small
over-correction for the excitation of the \cco\ line in cB58 (Baker et
al.\ \citeyear{bak04}), and the different adopted $\mu_{\rm L}$(CO)
and SFR for the Cosmic Eye (Coppin et al.\
\citeyear{cop07}).} These are by a factor of a few shorter than in
SMGs ($<$100\,Myr; e.g., Greve et al.\ \citeyear{gre05}) and
$>$30$\times$ shorter than in SFGs ($\sim$0.5--0.9\,Gyr; e.g., Daddi
et al.\ \citeyear{dad08}; Tacconi et al.\ \citeyear{tac10}). However,
SMGs have typically 10--50$\times$ higher SFRs than these LBGs, while
SFGs have comparable SFRs.

The ratio between $L_{\rm IR}$ ($\propto$SFR) and $L'_{\rm CO}$
($\propto$$M_{\rm gas}$) can be used as a measure of the star
formation efficiency. We find ratios of $\sim$260$\pm$150 and
$\sim$710$\pm$390 for cB58 and the Cosmic Eye, respectively,
comparable to what is found in nearby ULIRGs and SMGs (typically
$\sim$250, but with large scatter up to $>$1000; e.g., Tacconi et al.\
\citeyear{tac06}, \citeyear{tac08}), and substantially higher than the
ratios found in nearby spiral galaxies (typically $\sim$30--60; e.g.,
Gao \& Solomon \citeyear{gs04}). However, SMGs have $\sim$40$\times$
higher median $L'_{\rm CO}$ and $M_{\rm gas}$, and $\sim$3$\times$
broader CO lines than these LBGs (e.g., Coppin et al.\
\citeyear{cop08}).\footnote{Some SFGs have similarly narrow CO lines,
but likely just due to low disk inclinations (e.g., Daddi et al.\
\citeyear{dad10}).}

\subsection{Sizes of the Gas Reservoirs}

Detailed studies of nebular emission lines in the rest-frame
UV/optical suggest that cB58 and the Cosmic Eye have intrinsic sizes
of $\sim$1--2\,kpc (e.g., Seitz et al.\ \citeyear{sei98}; Stark et
al.\ \citeyear{sta08}), indicating that they are more compact than
SMGs and SFGs (e.g., Tacconi et al.\ \citeyear{tac08},
\citeyear{tac10}; Carilli et al.\ \citeyear{car10}; Daddi et al.\
\citeyear{dad10}).  The extent of the \aco\ emission in these
$z$$\sim$3 $L^{\star}_{\rm UV}$ LBGs is consistent with that of the
UV/optical light. It thus is unlikely that the gas is distributed on
scales as large as typical for the above types of galaxies.

\section{Conclusions}

We have detected luminous \aco\ emission toward the gravitationally
lensed $\sim$$L^{\star}_{\rm UV}$($z$$\simeq$3) LBGs cB58 ($z$=2.73)
and the Cosmic Eye ($z$=3.07). The ground-state CO line carries
30\%--40\% more luminosity than the previously detected \cco\ lines in
these galaxies. This implies that the $J$=3 lines are subthermally
excited. The gas masses, gas excitation, gas fractions, and star
formation efficiencies in these $z$$\sim$3 LBGs are consistent with
nearby luminous infrared galaxies, which also matches their observed
$L_{\rm IR}$. These LBGs have comparable SFRs to SFGs (Daddi et al.\
\citeyear{dad08}, \citeyear{dad10}; Tacconi et al.\ \citeyear{tac10}),
but their gas properties suggest that their star formation mode is
consistent with starbursts, rather than these high-$z$ disk galaxies
(which harbor comparatively long-lasting star formation at low
efficiencies; see Daddi et al.\ \citeyear{dad10b}; Genzel et al.\
\citeyear{gen10}). Even though the star formation {\em mode} and SSFRs
are consistent with SMGs at similar $z$, these LBGs are substantially
less massive, less extreme, less extended and (likely) less
dust-obscured systems. While SMGs may trace a brief, but common phase
in the evolution of massive galaxies, LBGs thus probably trace a
common phase in the formation of more `typical' (i.e.,
$\sim$$L^{\star}$) present-day galaxies (e.g., Somerville et al.\
\citeyear{som01}; Adelberger et al.\ \citeyear{ade05}; Conroy et al.\
\citeyear{con08}).

This consistent picture is obtained if one chooses a ULIRG-like
$\alpha_{\rm CO}$ conversion factor, and helps to motivate it. Both
cB58 and the Cosmic Eye are thought to have slightly sub-solar
metallicities ($Z$$\simeq$0.4 and 0.9\,\zsol; e.g., Baker et al.\
\citeyear{bak04}; Stark et al.\ \citeyear{sta08}), which may require
some modification to $\alpha_{\rm CO}$. However, as already discussed
by Baker et al.\ (\citeyear{bak04}), there is no consensus in the
literature on how severe the impact of metallicity on $\alpha_{\rm
CO}$ really is, in particular due to the fact that the CO lines arise
in optically thick gas. We thus do not modify $\alpha_{\rm CO}$ from
the canonical value for ULIRGs found by Downes \& Solomon
(\citeyear{ds98}), but do acknowledge the typical factor of a few
uncertainty inherent to this assumption (see also discussion by Coppin
et al.\ \citeyear{cop07}).

The observations presented here revise the masses and some of the
physical properties of the gas reservoirs in the lensed $z$$\sim$3
LBGs cB58 and the Cosmic Eye, highlighting the importance of \aco\
observations in comparatively `ordinary' (i.e., $\sim$$L^{\star}_{\rm
UV}$) high-$z$ galaxies. The gas reservoirs in both systems are
consistent with those in starburst regions of nearby luminous infrared
galaxies, providing supporting evidence that LBGs mark intense star
formation events in common, relatively low-mass galaxies at high $z$
(in comparison to SMGs). The conditions for star formation appear
markedly different from those in the massive, gas-rich star-forming
galaxies at high $z$ that were discovered recently (SFGs; Daddi et
al.\ \citeyear{dad08}, \citeyear{dad10}; Tacconi et al.\
\citeyear{tac10}), which are typically more gas-rich and massive, but
have lower star formation efficiencies.

The present investigation thus has identified differences between the
gas properties of differently selected, comparatively common high-$z$
star-forming galaxies that host less extreme star formation events
than SMGs and far-infrared-luminous high-$z$ quasars. With the rising
capabilities of the EVLA to study high-$z$ \aco\ emission in a more
unbiased manner, we thus are beginning to unravel the different
contributors to the gas mass and star formation histories of the
universe in a more direct way than possible so far. 

\acknowledgments 
We thank the referee for a helpful report. DR acknowledges support
from from NASA through Hubble Fellowship grant HST-HF-51235.01 awarded
by STScI, operated by AURA for NASA, under contract NAS\,5-26555. NRAO
is a facility of the NSF operated under cooperative agreement by AUI.

\end{document}